\documentclass[11pt, letter]{article}

\usepackage{stolenstyle}

\usepackage{amsmath,epsf,amssymb,latexsym,amsthm,setspace,bbm,array,pifont}

\usepackage[matrix,arrow,frame,import,curve,color]{xy}

\DeclareFontFamily{U}{rsf}{}
\DeclareFontShape{U}{rsf}{m}{n}{
  <5> <6> rsfs5 <7> <8> <9> rsfs7 <10-> rsfs10}{}
\DeclareMathAlphabet\Scr{U}{rsf}{m}{n}

\def\CO#1#2{{[#1,#2]}}
\def\AC#1#2{{\{#1,#2\}}}

\def\GUR{\GU(1)_{\text{R}}}

\def\C{{\mathbb C}}

\def\R{{\mathbb R}}
\def\Z{{\mathbb Z}}

\def\End{\operatorname{End}}

\def\Img{\operatorname{Im}}

\def\Rea{\operatorname{Re}}

\def\ch{\operatorname{ch}}

\def\tr{\operatorname{tr}}

\def\SU{\operatorname{SU}}
\def\GU{\operatorname{U{}}}

\def\Spin{\operatorname{Spin}}

\def\GE{\operatorname{E}}

\def\Lsl{\operatorname{\mathfrak{sl}}}

\def\p{\partial}
\def\pb{\bar{\partial}}

\def\ff#1#2{{\textstyle\frac{#1}{#2}}}

\def\cA{{\cal A}}

\def\cE{{\cal E}}
\def\cF{{\cal F}}

\def\cP{{\cal P}}

\def\cS{{\cal S}}

\def\ep{{\epsilon}}

\newcommand\thetab{\overline{\theta}}

\newcommand\lambdab{\overline{\lambda}}
\newcommand\mub{\overline{\mu}}
\newcommand\nub{\overline{\nu}}
\newcommand\xib{\overline{\xi}}

\newcommand\vphi{\varphi}

\newcommand\Thetab{\overline{\Theta}}

\newcommand\Sigmab{\overline{\Sigma}}

\newcommand\Omegab{\overline{\Omega}}

\newcommand\gh{\widehat{g}}

\newcommand\cb{\overline{c}}

\newcommand\jb{\overline{\jmath}}
\newcommand\kb{\overline{k}}

\newcommand\nb{\overline{n}}

\newcommand\pbar{\overline{p}} 

\newcommand\xb{\overline{x}}

\newcommand\zb{\overline{z}}

\newcommand\Rh{\widehat{R}}

\newcommand\Ab{\overline{A}}

\newcommand\Fb{\overline{F}}

\newcommand\Tb{\overline{T}}

\def\bSigma{{\boldsymbol{\Sigma}}}
\def\bTheta{{\boldsymbol{\Theta}}}

\def\bF{{\boldsymbol{F}}}
\def\bT{{\boldsymbol{T}}}

\def\bg{{\boldsymbol{g}}}

\def\ba{{\mathbf{a}}}

\def\bFb{{\overline{\bF}}}

\def\preprint{ {{EFI-14-5}}\\{{IPhT-T14/026}} \\ {{MIFPA-14-07}} }

\title{Heterotic fluxes and supersymmetry}
\author[a] {Ilarion V.~Melnikov,}
\author[b] {Ruben Minasian,}
\author[\,c] {and Savdeep Sethi}
\affiliation[a]{George P. and Cynthia W. Mitchell Institute for Fundamental Physics and Astronomy,\\
Texas A\&M University, College Station, TX 77843, USA}
\affiliation[b]{ Institut de Physique Th{\'e}orique, CEA/Saclay, 91191 Gif-sur-Yvette Cedex, France}
\affiliation[c]{Enrico Fermi Institute, University of Chicago, Chicago, IL 60637, USA}
\emailAdd{ilarion@phys.tamu.edu}
\emailAdd{ruben.minasian@cea.fr}
\emailAdd{sethi@uchicago.edu}
\abstract{We show that the formal $\alpha'$ expansion for heterotic flux vacua is only sensible when flux quantization and the appearance of string scale cycles in the geometry are carefully taken into account.  We summarize a number of properties of solutions with N=1 and N=2 space-time supersymmetry.}

\begin{document}

\maketitle

\section{Introduction} \label{s:intro}
A heterotic flux vacuum is a solution of the classical heterotic string equations of motion with a non-trivial three-form field strength $H$.  Solutions of this sort have a great deal to teach us about the role of fluxes in stringy geometry and the landscape of string vacua, yet despite being introduced very soon after the discovery of the heterotic string, there remain many basic questions about their properties.  For instance, even in cases where a dual IIA or F-theory perspective might make it clear that a particular solution with flux should exist, understanding the existence directly in the heterotic string requires a careful analysis of the $\alpha'$ expansion, and naive considerations can easily lead to incorrect interpretation and results.  The aim of this note is to explain some of these subtleties in a number of examples.  Our hope is that the lessons learned will be of use in charting the heterotic landscape.

We will be interested in perturbative heterotic compactifications that preserve super-Poincar\'e invariance on $\R^{1,3}$ with $4$ or $8$ supercharges.  In this case both the world-sheet theory, with its (0,2) superconformal invariance, and the effective space-time SUSY theory act to constrain possible mechanisms that might destabilize a solution.  We will also consider SCFTs that can be described by a heterotic geometry, in other words a (0,2) SUSY non-linear sigma model (NLSM).  While a NLSM Lagrangian description is really only applicable in a large radius limit, where the curvatures of all space-time fields can be made small, we will also apply our considerations to geometries with string-scale cycles.  While this might seem a priori a formal exercise, experience has shown it to be useful. Most of our examples of heterotic flux vacua were found using space-time dualities applied to large volume compactifications. Although a duality transformation can generate string scale structures, by definition it also preserves the feature that the new background also solves the space-time equations of motion. This is something we should be able to verify explicitly.  Our basic concern is therefore to correctly identify the geometric ingredients necessary to obtain a formal solution to leading order in $\alpha'$ to the supersymmetry conditions and the Bianchi identity.  

There are two key lessons we learn from this exercise.  First, one has to simultaneously consider the $\alpha'$ expansion for both the solution and the equations of motion.  Once this is done, the leading terms in the $\alpha'$ expansion can be treated self-consistently.  
 Second, it is not physically correct to treat the heterotic space-time equations of motion, truncated to include just the leading order $\alpha'$ corrections, as a closed system.  One can certainly study solutions to this system of equations as a mathematics problem; however, such an analysis misses essential changes to the system of equations from interactions that are higher order in $\alpha'$. We will see this explicitly in specific examples. 
 
These are special situations where treating the truncated system as an exact system works better than one might expect, but those cases typically involve models that preserve extended rather than minimal space-time supersymmetry.
While perturbative solutions to the space-time equations of motion are useful guides to some properties of string compactifications, we should stress that developing world-sheet techniques is indispensable if we are to properly understand the conformal field theories that define these string vacua. 

Before we tackle the details, it is useful to catalogue the known heterotic geometries of this type.  By far the most studied and well-understood class is described by picking a stable holomorphic bundle $\cE$ over  a Calabi-Yau $3$-fold $X$.  These solutions have a large radius limit in the moduli space, where the flux $H$ remains of order $\alpha'$ as we take the limit.  The classic work of~\cite{Witten:1986kg} described how to incorporate $O(\alpha')$ corrections to solve at least some of the equations of motion at $O(\alpha')$.  In this work we will show that the leading order Calabi-Yau and bundle geometry can be perturbed to satisfy all of the supersymmetry conditions and the heterotic Bianchi identity at $O(\alpha')$.

Another class of examples is offered by topologically non-K\"ahler heterotic compactifications.  It is reasonable to expect that such compactifications will describe the generic heterotic geometry,\footnote{Dual perspectives also suggest that there is a larger landscape of non-geometric heterotic flux vacua~\cite{McOrist:2010jw}, so the geometric setting is by no means exhaustive.} but we lack techniques to build large classes of examples of such spaces.  See, however,~\cite{Melnikov:2012nm, Heckman:2013sfa} for very recent progress.  The basic technical challenge comes from having to work with the (comparatively poorly understood) complex geometry, as opposed to the more familiar K\"ahler setting.  It is also important to keep in mind that such constructions come with an important caveat:  the presence of H-flux means that the solutions necessarily have string-scale cycles, and the world-sheet NLSM is necessarily strongly coupled.  Given the dual F/M-theory constructions of such vacua we can be reasonably confident that the vacua exist~\cite{Dasgupta:1999ss}, but it is not a priori clear that the heterotic geometric description based on an $\alpha'$ expansion will give sensible results.  We will show that the $O(\alpha')$ ``corrections'' can be incorporated in a self-consistent fashion, provided that we keep track of $\alpha'$ factors in the solution as well as in the equations.

The plan of the rest of the note is as follows: we will warm up by examining the $O(\alpha')$ corrections to the SUSY equations for heterotic vacua with a large radius limit.  We will then tackle the same issue in known heterotic flux vacua, making a distinction between configurations that preserve N=1 and N=2 space-time SUSY.  We will wrap up with some concluding remarks. 

\acknowledgments IVM would like to thank CEA Saclay and the Enrico Fermi Institute for hospitality while this work was being completed;  RM would like to thank the Mitchell Institute for hospitality. IVM and RM thank S.~Theisen for discussions and collaboration in the early stages of this project, and we thank K.~Becker and L.-S.~Tseng for useful discussions.  The three of us would like to thank the organizers and participants of the String Geometry and Beyond 2013 workshop in Costa Rica and the Soltis Center for an inspiring environment.   This work is supported in part by the NSF Focused Research Grant DMS-1159404  and Texas A\&M (IVM), Agence Nationale de la Recherche under grant 12-BS05-003-01 (RM), and NSF Grant No. PHY-1316960 (SS).

\section{The Calabi-Yau warm-up} \label{s:CY}

\subsection{Conditions for supersymmetry}
Consider a heterotic compactification that preserves N=1 d=4 super-Poincar{\'e} invariance.  The supersymmetry conditions are known since the seminal work of~\cite{Strominger:1986uh} and can be stated in the language of $G$-structures~\cite{Gauntlett:2003cy}.  This can also be seen as a consequence of one-loop Weyl invariance of the world-sheet theory~\cite{Hull:1986kz,Nibbelink:2012wb}.  As this is familiar, we simply state the conditions.  The target space $X$ must be a complex manifold with $\SU(3)$ structure and trivial canonical bundle.  The Hermitian form $\omega \in \Omega^2(X)$ and the (3,0) holomorphic form in $\Omega \in \Omega^3(X)$ must have the same exact Lee form $\beta \in \Omega^{1}(X)$, and this determines both the gauge-invariant torsion $H \in \Omega^3(X)$, as well as  the dilaton $\vphi$ up to a constant.  The heterotic gauge bundle is in general some principal $G$-bundle $\cP$ with $G \subset \GE_8 \times \GE_8$ or $G\subset \Spin(32)/\Z_2$, and the corresponding curvature $\cF$ satisfies zero-slope Hermitian Yang-Mills (HYM) equations.

In terms of the $\alpha'$ expansion the conditions are, up to $O(\alpha'^2)$ corrections~\cite{Bergshoeff:1989de},
\begin{align}
\label{eq:SUSY}
\omega^3  & = \ff{3i}{4} \Omega \Omegab~,& 
\omega \Omega & = 0~,&
\Omega^2 & = 0~,\nonumber\\
d(e^{-\vphi} \omega^2) & = 0~, &
d(e^{-\vphi} \Omega )   & = 0~, &
H & = i(\pb -\p) \omega~,&\nonumber\\
\cF \Omega & = 0~,&
\cF \Omegab & = 0~,&
\omega^2 \cF & = 0~.
\end{align}
Note that we suppress the $\wedge$ in these form equations.
In addition we need to satisfy the Bianchi identity
\begin{align}
\label{eq:BIANCHI}
d H = 2i \p\pb \omega = \frac{\alpha'}{4} \left( \tr R_+^2 - \tr \cF^2 \right) + O(\alpha'^2)~.
\end{align}
Here $R_+$ is the curvature of the $H$-twisted connection on the tangent bundle.  In terms of a real coordinate basis we can write the twisted connection 1-form  $\cS_{+}$ in terms of the Levi-Civita connection $\Gamma$ and $H$:
\begin{align}
\label{eq:plustor}
 dx^B \cS^{A}_{+BC}   = 
 dx^B (\Gamma^{A}_{~BC} + \frac{1}{2} H^{A}_{~BC})~.
\end{align}
The trace of the gauge bundle term is written a bit imprecisely, but the normalization is easily fixed in terms of the index of embedding describing $G$ as a subgroup of $\GE_8\times\GE_8$ or $\Spin(32)/\Z_2$.  The Bianchi identity of leads to the famous Green-Schwarz cancellation condition equating the Pontryagin classes $p_1(T_X)$ and an appropriate multiple, determined by the index of the embedding, of $p_1(\cP)$.  This topological condition is independent of a particular choice of connections in~(\ref{eq:BIANCHI}); however, it is important to keep in mind that to find SUSY vacua to this order in $\alpha'$ we must use the indicated connections.  

\subsection{Leading corrections to Calabi-Yau geometry}

Consider the text-book example of a space-time SUSY heterotic compactification:  a Calabi-Yau $3$-fold $X$ with $(\omega_0,\Omega_0)$ defining a K\"ahler Ricci-flat metric, and a stable holomorphic vector bundle $\cE \to X$ satisfying the anomaly cancellation condition, which now simplifies to $\ch_2(\cE) = \ch_2(T_X)$.  Let $\cA_0$ be corresponding HYM connection with curvature $\cF_0$.  For simplicity we will take $c_1(\cE) = 0$ and assume that $\cE$ is irreducible.   Clearly this data, together with a  constant dilaton $\vphi=\vphi_0$, solves~(\ref{eq:SUSY}) and~(\ref{eq:BIANCHI}) at $O(\alpha'^0)$.  As we will now show, we can suitably modify $\omega$, $\Omega$ and $\cA$ so as to solve all of the conditions at $O(\alpha')$.\footnote{For this to be sensible we need to know the equations at $O(\alpha')$.  Thanks to~\cite{Bergshoeff:1989de} we know the ten-dimensional SUSY variations up to $O(\alpha'^2)$ terms, and on six-dimensional bosonic backgrounds they are given by~(\ref{eq:SUSY}) and~(\ref{eq:BIANCHI}).}  Specifically, we make an Ansatz
\begin{align}
\omega &= \omega_0 + \alpha' \left[ \ff{4}{3} \vphi_1 \omega_0 + \eta \right]~,&
\Omega & = (1+ 2\alpha' \vphi_1) \Omega_0~,&
\vphi & = \vphi_0 + \alpha' \vphi_1~,&
\cA & = \cA_0 + \alpha' \cA_1~.
\end{align}
Plugging these into our equations, we are left with
\begin{align}
\label{eq:CY}
\omega_0^2 \eta &= 0~,\nonumber\\
\omega_0 d \left[ \ff{1}{3}\vphi_1 \omega_0 + \eta\right] & = 0~, \nonumber\\
i\p\pb \left[ \ff{4 }{3}\vphi_1 \omega_0 + \eta\right] & = \ff{1}{8} \left[ \tr R_0^2 - \tr \cF_0^2\right] = i \p\pb \sigma~,\nonumber\\
\omega_0^2 \pb \cA_1 & = -2 \omega_0 \eta \cF_0~.
\end{align}
The first equation, a left-over of the $\SU(3)$ structure requirements, says that $\eta$ is a primitive (1,1) form; the second ensures that $\omega$ remains conformally balanced to this order in $\alpha'$.  To derive the third equation we used $\omega$ to obtain $H$ and plugged it into the Bianchi identity; furthermore, since $R_0$ and $\cF_0$ are both (1,1) forms, the right-hand side of the Bianchi identity is both $d$-exact and pure (2,2), and hence we can use the $\p\pb$-lemma on $X$ to write it as $i\p\pb \sigma$ for some real (1,1) form $\sigma$.

\subsubsection*{A little Lefschetz decomposition}
We will show that~(\ref{eq:CY}) have solutions for $\vphi_1$, $\eta$ and $\cA_1$. 
The solution depends on properties of Laplacians for the K\"ahler structure $(\omega_0,\Omega_0)$ on $X$, and while it can be obtained by wading through a mire of indices, we prefer to use some basic facts about the Lefschetz decomposition to simplify the equations and find solutions. To that end we now review some of those techniques.  More details can be found in~\cite{Griffiths:1978pa}.  

A compact complex $n$-fold $X$ with K\"ahler structure $(\omega_0,\Omega_0)$ can be equipped with the differentials $\p,\pb$ and $d = \p+\pb$.  Constructing adjoint operators we obtain Laplacians $\Delta_{\p}$, $\Delta_{\pb}$, and $\Delta_{d}$ --- e.g. $\Delta_{\p} = \p\p^\dag +\p^\dag\p$.  The forms $\Omega^k(X) = \bigoplus_{p+q =k} \Omega^{p,q}(X)$ can be decomposed according to the $\Lsl_2\C$ algebra generated by 
\begin{align}
L &: \Omega^k(X) \to \Omega^{k+2}(X),~\nonumber\\
L &: \tau \mapsto \omega_0 \tau~
\end{align}
and its adjoint $L^\dag : \Omega^{k}(X) \to \Omega^{k-2}(X)$, which satisfy
\begin{align}
\CO{L^\dag}{L}  \Omega^{p,q}(X) = (n-p-q) \Omega^{p,q}(X)~.
\end{align}
We have isomorphisms
\begin{align}
L^k & :\Omega^{n-k} \to \Omega^{n+k}~, &
\ker L^{k+1} \cap \Omega^{n-k} \simeq \ker L^\dag \cap \Omega^{n-k}~.
\end{align}
These relations can be refined to $\Omega^{p,q}$ in the obvious fashion and allow us to define primitive forms as those annihilated by $L^\dag$ or, equivalently, by an appropriate power of $L$.
This decomposition descends to cohomology, giving us a notion of primitive classes. 
Another  key property is that $L$ ($L^\dag$) commutes with $\p,\pb$ ($\p^\dag,\pb^\dag$), while
\begin{align}
\CO{{L^\dag}}{\pb} &= -i \p^\dag~,&
\CO{{L^\dag}}{\p} &= i \pb^\dag~,&
\CO{L}{\pb^\dag} &= -i \p~,&
\CO{L}{\p^\dag} &= i\pb~.
\end{align}
As a consequence, we obtain $\Delta_{\p} = \Delta_{\pb} = \ff{1}{2} \Delta_{d}$, $\AC{\p}{\pb^\dag} = 0$, and the $\p\pb$--lemma.\footnote{A complex manifold with a $\p\pb$-lemma need not be K\"ahler, but every K\"ahler manifold has a $\p\pb$-lemma.}  Another useful result is
\begin{align}
\label{eq:Lddbar}
\CO{L^\dag}{\p\pb} & = i \pb^\dag \pb - i \p\p^\dag~,\nonumber\\
\CO{(L^\dag)^2}{\p\pb} & = 2i (\pb^\dag \pb -\p\p^\dag) L^\dag + 2 \pb^\dag \p^\dag~.
\end{align}
We will need one more simple consequence of this structure:  any real (1,1) form $\sigma$ may be decomposed as $\sigma = h\omega_0 + \kappa$, where $\kappa$ is a real (1,1) form satisfying $\omega_0 \p\pb \kappa = 0$.  To prove this, we simply need to solve $\omega_0 \p\pb \sigma = \p\pb h \omega_0^2$ for a real function $h$ and set $\kappa = \sigma-h\omega_0$.  We have
\begin{align}
\omega_0 \p\pb \sigma - \p\pb h \omega_0^2 & = 0 &\iff&&  (\L^\dag)^3 (\omega_0 \p\pb \sigma - \p\pb h \omega_0^2 )& = 0~.
\end{align}
Using the commutators of $L^\dag$, $L$ and $ \p\pb$ just described, we find this is equivalent to
\begin{align}
\Delta_{\pb} h = \ff{1}{2} \Delta_{\pb} L^\dag \sigma - \ff{i}{2} \pb^\dag \p^\dag \beta~.
\end{align}
Since the right-hand side is orthogonal to all harmonic forms there exists a requisite $h$.

\subsubsection*{Solution to $O(\alpha')$}
We now have the tools to show that~(\ref{eq:CY}) has solutions.  Using the last result from the previous section, we decompose the source in the Bianchi identity as $\sigma = h \omega_0+ \kappa$ with $\p\pb\kappa \omega_0 = 0$.  We can take both $h$ and $\kappa$ to be orthogonal to all harmonic forms.  We can solve the Bianchi equation by setting
\begin{align}
\eta = \sigma - \ff{4}{3} L \vphi_1 + \p \xi + \pb\xib + i \p\pb f~,
\end{align}
where $f$ is any real function and $\xi$ any (1,0) form.  Setting aside for a moment the HYM equation, we then need to solve
\begin{align}
L^\dag ( \sigma - \ff{4}{3} L \vphi_1 -\pb \xi - \p \xib +i \p\pb f) & = 0~, \nonumber\\
L d( L h + \kappa - L \vphi_1 -\pb \xi - \p \xib) & = 0~.
\end{align}
The first of these is a Laplace equation for $f$ with a source orthogonal to harmonic functions, and it has a solution $f$ for any $\xi$ and $\vphi_1$.  The second equation is a 5-form equation that can be decomposed into complex conjugate (3,2) and (2,3) components.  Concentrating on the first of these, and setting $\vphi_1 = h$, we need to solve
\begin{align}
L(\p\kappa - \p\pb \xi) = 0
\end{align}
for $\xi$ for a $\kappa$ satisfying $L \p\pb \kappa = 0$.  Applying $(L^\dag)^2$ we have
\begin{align}
0 = (L^\dag)^2 L (\p\kappa-\p\pb\xi) = 2 L^\dag (\p\kappa - \p\pb \xi) \implies i \pb^\dag \kappa + \p (L^\dag\kappa) - i (\pb^\dag\pb - \p\p^\dag) \xi~.
\end{align}
We can simplify this by observing that $\pb^\dag \xi = 0$, so that 
\begin{align}
(\pb^\dag\pb -\p\p^\dag ) \xi = (\Delta_{\pb} -\p\p^\dag)\xi = \p^\dag \p \xi~.
\end{align}
Thus, we need to find $\xi$ satisfying
\begin{align}
\label{eq:xifin}
\p^\dag\p \xi = (\pb^\dag \kappa - i \p L^\dag \kappa)~.
\end{align}
Since $L \p\pb \kappa = 0$ is equivalent to $\p^\dag(\pb^\dag \kappa -i \p L^\dag\kappa) = 0$, the (1,0)  form on the right-hand side is $\p^\dag$-closed, and hence on our Calabi-Yau is $\p^\dag$-exact.  But, any $\p^\dag$-exact (1,0) form can be written as $\p^\dag \p \tau$, and setting $\xi = \tau$ we solve~(\ref{eq:xifin}).

The remaining supersymmetry condition is the correction to HYM --- the last equation in~(\ref{eq:CY}).  To see that there is a solution, we first note that for an irreducible bundle $H^{0,0}_{\pb} (X,\End \cE) = 0$, where $\End \cE \subset \cE \otimes \cE^\ast$ denote the traceless endomorphisms.  Since $\End \cE \simeq \End \cE^\ast$, Poincar{\'e} duality implies $H^{3,3}_{\pb} (X,\End \cE) = H^{0,0} (X,\End \cE^\ast) = 0$.  Therefore, $\omega_0\eta \cF_0 = \pb \pi$ for some $\pi \in \Omega^{3,2}(X,\End\cE)$.  Since $L^2 :\Omega^{1,0}(X,\End \cE) \to \Omega^{3,2}(X,\End \cE)$ is an isomorphism, any such $\pi$ can be written as $\pi = \omega_0^2 \mu$ for a $\mu \in \Omega^{1,0}(X,\End \cE)$, so that the linearized HYM equation becomes
\begin{align}
\omega_0^2 \pb \cA_1 = \omega_0^2 (-2 \pb \mu)~.
\end{align}
Clearly $\cA_1 = -2 \mu$ yields the desired solution.

We have shown that, as expected,  a large radius Calabi-Yau heterotic compactification can be deformed to obtain a solution to the SUSY conditions and Bianchi identity to $O(\alpha')$.  It turns out that one can do more~\cite{Li:2004hx,Andreas:2010cv}.  Suppose we consider a different system of equations where the SUSY conditions~(\ref{eq:SUSY}) and the Bianchi identity are assumed to hold without any additional $\alpha'$ corrections.  It can then be shown that the full system has a solution provided the curvature $R_+$ satisfies the subsidiary condition of being HYM.  This subsidiary condition can be motivated by the result~\cite{Ivanov:2009rh} that the $O(\alpha')$ solution to SUSY and Bianchi conditions is a solution of the heterotic equations of motion if and only if $R_+$ obeys HYM up to $O(\alpha')$. 

It is not clear what one is to make of this interesting result.   While it is possible, as suggested in~\cite{delaOssa:2014cia}, that this structure persists at $O(\alpha'^2)$, the explicit $\alpha'^3$ corrections to the equations of motion~\cite{Grisaru:1986dk,Gross:1986iv} already known for heterotic vacua with (2,2) world-sheet supersymmetry certainly violate HYM for the tangent bundle curvature. This is an important point that we should stress: aside from exceptional cases, it is not physically sensible to demand an exact solution to the truncated heterotic space-time equations of motion.

\section{Heterotic flux solutions} \label{s:n2sol}
We now turn to the flux vacua without large radius limit.  In this case the geometry is rather more involved:  not only constructions of admissible topologies few and far between, but we also lack the analogues of Yau's and Donaldson-Uhlenbeck-Yau's theorems, which in the Calabi-Yau case yield necessary and sufficient conditions on complex structure and K\"ahler class in order to have a solution of the leading order supergravity equations of motion.  It would be extremely useful to have such conditions, but without them we must proceed by examining a class of examples.

Essentially all known heterotic flux vacua are variations on a single theme inspired by a dual F/M-theory compactification on $\text{K3}\times\text{K3}$~\cite{Dasgupta:1999ss}.  There have been a number of generalizations, including attempts to check that the equations of motion are satisfied; see, for example,~\cite{Becker:2002sx, LopesCardoso:2002hd, Becker:2008rc, Becker:2009df, Andriot:2009fp, Becker:2009zx, Carlevaro:2011mn}. For simplicity we will focus on a particularly simple configuration and comment on the additional bells and whistles below.

\subsection{Review of geometric set-up}
The target space $X$ is a principal $T^2$ bundle $\pi : X \to M$  over a K3 base $M$ with each fiber a square torus of fixed area.   The topology and complex structure of such $X$ is nicely described in~\cite{Goldstein:2002pg}.  The topology of $X$ is fixed by a choice of two linearly independent nowhere vanishing real $1$-forms $\bTheta^{I} \in \Omega^1(X)$ dual to the vertical vector fields $\frac{\p}{\p\theta^{I}}$, $I=1,2$.  In general these are not closed but satisfy 
\begin{align}
d\bTheta^{1,2} &= \pi^\ast (\bF^{1,2})~,&
\bF^{1,2} & \in H^2(M,2 \pi \Z)~.
\end{align}
In what follows we will leave off the explicit pull-back $\pi^\ast$ unless it is likely to cause confusion.  Unless otherwise noted, we will also assume that the $\bF^I$ are linearly independent in $H^2(M,2\pi \Z)$.

The complex structure of $X$ is determined by a choice of complex structure on $M$ compatible with a complex fiber.  Let $U \in M$ be an open neighborhood with complex coordinates $z^i$, $i=1,2$. The corresponding local structure of $X$ is given by $U\times T^2$, where $\theta^I \sim \theta^I + 2\pi$ are coordinates on $T^2$.  We write $\bTheta^I = d\theta^I + A^I$, where the gauge-fields $A^I$ are horizontal 1-forms on $U$.  We can take $\theta = \theta^1 + i \theta^2$ as the holomorphic coordinate on the square $T^2$ fiber.\footnote{More generally we could set $\theta_\alpha = \theta^1_\alpha + \tau \theta^2_\alpha$.  The case of constant $\tau$ is a simple modification, while $\tau$ varying holomorphically over the base is a bit more subtle. See~\cite{Becker:2009df}\ for a discussion. }   This gives an almost complex structure on $X$, with (1,0) forms $\Theta = \bTheta^1 + i \bTheta^2$ and $dz^i$.  This is integrable if and only if $d\Theta$ has no (0,2) component its decomposition.  From above we see that if we set $\bF = \bF^1 + i \bF^2$, we need
\begin{align}
\bF &= F+F'~,&
F & \in H^{1,1}(M)~,&
F' & \in H^{2,0}(M)~.
\end{align}
The complex conjugate forms satisfy $\bF^1 -i \bF^2 = \Fb + \Fb'$, with $\Fb \in H^{1,1}(M)$ and $\Fb' \in H^{0,2}(M)$.  
It is useful to write the $F$ and $F'$  in terms of various components of the $\bF^I$:
\begin{align}
\label{eq:recurv}
F &= \bF^{1,1}_1 + i \bF^{1,1}_2~,&
F' &= 2 \bF^{2,0}_1~, &
\Fb & = \bF^{1,1}_1 -i \bF^{1,1}_2~,&
\Fb' & = 2 \bF^{0,2}_1~.
\end{align}
We used $(\bF^{1} + i \bF^{2})^{0,2} = 0$.

The SUSY conditions of~(\ref{eq:SUSY}) are satisfied in a straightforward manner.  Let $(\omega_0,\Omega_0,\cA)$ denote a Calabi-Yau structure on the K3 base $M$ supplemented by a HYM connection $\cA$ for a stable holomorphic bundle $\cE' \to M$.  In particular, $\omega_0$ and $\Omega_0$ are closed and satisfy the usual $\SU(2)$ structure relations 
\begin{align}
\omega_0^2 &= \ff{1}{2} \Omega_0 \Omegab_0~,& \omega_0 \Omega_0 &= \Omega_0^2  = 0~.
\end{align}
Suppose we choose the complex structure on $M$ such that $\bF = \bF^1 + i \bF^2$ satisfies
\begin{align}
\omega_0 \bF &= 0~ &\Omega_0 \bF & = 0~.
\end{align}
The second condition is just that $\bF$ has no (0,2) component.
In that case, denoting the constant area of the $T^2$ fiber by $\ba$, we find that
\begin{align}
\label{eq:fluxSUSY}
\omega_X &= e^{2\vphi} \omega_0 +\frac{ i \ba}{2} \Theta\Thetab~,&
\Omega_X & = e^{2\vphi} \sqrt{\ba} \Omega_0 \Theta~,&
\cF & = \pi^\ast \pb \cA~
\end{align}
satisfy~(\ref{eq:SUSY}) on $X$ with bundle $\cE = \pi^\ast \cE'$.   The torsion is determined to be
\begin{align}
\label{eq:fluxtor}
H & = i (\pb -\p) \omega_X  = i \omega_0 (\pb-\p) e^{2\vphi} + \frac{\ba}{2} (\Fb'-\Fb) \Theta + \frac{\ba}{2} (F'-F)\Thetab~\nonumber\\
&  = H_{\text{hor}} + H_I \bTheta^I = H_{\text{hor}} + \ba(\bF_I^{2,0} + \bF_I^{0,2} - \bF^{1,1}_I) \bTheta^I~.
\end{align}
Finally, we have the Bianchi identity
\begin{align}
\label{eq:fluxBianchi}
d H_{\text{hor}} + \ba \left[ F'\Fb' -F\Fb\right] + \ba \p\Fb' \Theta +\ba \pb F' \Thetab = \frac{\alpha'}{4} \left[ \tr R_+^2 -\tr \cF^2\right] + O(\alpha'^2).
\end{align}

\subsection{Comments on flux solutions}
Having presented the basic geometry, we are ready to describe some of the properties of these solutions.  This will help us to frame a discussion of some of their features and subtleties.

\subsubsection*{$T^2\times {K3}$ compactification}
The reader who went through the Calabi-Yau 3-fold warm-up above may have wondered what happens if we ask the same sort of question of K3 compactification:  can we perturb the $O(\alpha'^0)$ solution to obtain an $O(\alpha')$ result?  Indeed we can.  Consider the trivial fibration $T^2\times\text{K3}$ obtained by setting $\bF^I = 0$.  In that case, we just need to solve~(\ref{eq:fluxBianchi}) for the dilaton profile.   Using the same sort of reasoning as in the 3-fold case, we can write the source as
\begin{align}
\tr R_+^2 -\tr \cF^2 = i\p\pb f \omega_0
\end{align}
for some real function $f$ and solve~(\ref{eq:fluxBianchi}) to $O(\alpha')$ by setting $\vphi = \alpha' f/16 + O(\alpha'^2)$.
In this case the perturbed solution is much less involved than in the 3-fold case:  the $\alpha'$-corrected metric on $M$ remains conformally Calabi-Yau.

\subsubsection*{N=2 versus N=1 solutions}
A generic choice of the $\bF^I$ fluxes will have $F' \neq 0$ and only preserve 4 supercharges in space-time.  However, we can also make a less generic choice, where $\bF^I \in H^{1,1}(M)$ and $\omega_0 \bF^I =0$, i.e. $\bF^{I}$ are both HYM. In this case the SUSY equations are actually symmetric under $\SU(2)$ rotations of $\omega_0, \Rea \Omega_0, \Img \Omega_0$, and this $\SU(2)$ signals the preservation of $8$ supercharges in space-time.

This geometric set-up is natural, in fact required, in the context of heterotic compactification with 8 supercharges:  if the requisite world-sheet supersymmetries~\cite{Banks:1988yz} are realized geometrically, then $X$ must be a principal $T^2$ bundle of a K3 surface  $M$ with a fiber of constant area, $B$-field and complex structure~\cite{Melnikov:2010pq,Melnikov:2012cv}.

On the other hand, the class of solutions with just 4 supercharges obtained by taking fluxes with $F' \neq 0$ are rather special in the landscape of heterotic vacua.  For instance, an elliptically fibered Calabi-Yau three-fold also has a fibration structure, but it is neither principal nor over a K3.  Still, the principal $T^2$ bundles are certainly an interesting distinguished class.  One may hope that their relation to backgrounds with 8 supercharges might make them more amenable to analysis.

\subsubsection*{Consequences of flux quantization}
It is an oft-repeated stringy mantra that flux quantization fixes parameters of supergravity solutions.  While this is familiar in the case of type II fluxes valued in appropriate de Rham cohomology groups, it is not so obvious for heterotic torsion since in general $d H \neq 0$.  By decomposing the flux into horizontal and vertical components and a judicious reparametrization of space-time Lorentz, gauge, and B-field gauge transformations, it is possible to identify quantization conditions that are obeyed by the $H_I$ --- the 2-forms appearing in the horizontal--vertical decomposition of $H$~\cite{Melnikov:2012cv}:
\begin{align}
H_I \in H^2(M, 2\pi\alpha' \Z)~.
\end{align}
This has a key consequence for the solution:  the ``parameter'' $\ba$ --- the area of the fiber $T^2$--- is in fact quantized in units of $\alpha'$.\footnote{In general the torus need not be at a point of enhanced symmetry.  When it is, a fibered WZW model of the sort discussed in~\cite{Adams:2009av,Melnikov:2012cv}\ can provide a more appropriate description.}

One can also show that there is a similar quantization condition on the $B$-field carried by the $T^2$ fiber.  The quantization conditions become a little bit more subtle in the presence of Wilson lines for the gauge bundle, but that complication can be addressed.  More details are given in section~2.5 of~\cite{Melnikov:2012cv}.

\subsubsection*{T-duality orbits and lift to $8$ dimensions}
While the area of the $T^2$ fiber is fixed, we do have the freedom to take the base K3 to be arbitrarily large.  In this large radius K3 limit we recover an $8$-dimensional theory --- the compactification on $T^2$.  While this is obviously not a large radius limit, it still offers a useful way of thinking about the theory.  In particular we can look for a possible dual F-theory description in terms of a fibered $8$-dimensional duality.

The $8$-dimensional perspective makes it apparent that there is a large redundancy in the nominal classification of solutions --- this is simply a consequence of the $O(2,18)$ duality of the $T^2$-compactification:  there are $O(2,18)$ transformations that allow us to exchange the ``physical'' $\GU(1)$s associated to the Kaluza-Klein reduced metric and $B$-field with the ``gauge'' $\GU(1)$s.  Such T-duality relations have been explored in~\cite{Evslin:2008zm, Israel:2013hna}.
This does not mean that every compactification with a non-trivial $T^2$-fibration is on a $T$-duality orbit of a $T^2\times \text{K3}$ compactification.  Clearly an N=1 vacuum cannot be obtained in this fashion, but the statement holds for N=2 vacua as well:  while $T^2 \times \text{K3}$ vacua necessarily have  a gauge group\footnote{We do not count the graviphoton here.} with rank $r \ge 3$, we can find examples of fibered solutions with $r < 3$.

Another property that is made clear from the $8$-dimensional point of view is the correlation between $\GU(1)$ space-time gauge symmetries and global world-sheet symmetries of the $T^2$ NLSM.  If the fibration is trivial then there are four Kac-Moody (KM) algebras associated to the heterotic string on $T^2$:  two come from the left, non-supersymmetric side of the string, while two more arise from the supersymmetric side.  The latter give rise to the graviphoton in the gravity multiplet and the gauge boson in the axio-dilaton vector multiplet in $4$ dimensions.  Turning on a non-trivial fibration breaks at least some of these symmetries.  Examination of the NLSM Lagrangian shows that $\ba F_I - H_I \neq 0$ breaks the left-moving chiral currents, while $\ba F_I + H_I \neq 0$ breaks the right-moving chiral currents.  From~(\ref{eq:recurv}) we then see
\begin{align}
\label{eq:brokenKM}
\text{{broken left symmetries}} :~~ & \ba\bF_I - H_I = 2\ba \bF_I^{1,1} \neq 0~, \nonumber\\
\text{{broken right symmetries}}:~~ & \ba\bF_I + H_I = 2\ba (\bF_I^{2,0}+\bF_I^{0,2}) \neq 0~.
\end{align}
This fits perfectly with the SUSY variations of the $d=8$ supergravity theory and underscores the different roles played by the (1,1) and (2,0) components of $\bF$:  the former is associated to breaking left-moving symmetries that are very much like the remaining heterotic gauge symmetries, while the latter is invariably tied to the gravity sector associated to the supersymmetric side of the string.

\section{Torsional connection and the Bianchi identity} \label{s:hetbianchi}
The reader has no doubt noticed that while we have discussed many aspects of the known heterotic flux solutions, we have stopped short of showing that the Bianchi identity holds to $O(\alpha')$.  It was observed already in~\cite{Strominger:1986uh} that solving~(\ref{eq:BIANCHI}) can be relatively easy or difficult, depending on which connection is used to compute the $\tr R^2$ term.  Given any connection, say the Levi-Civita $\Gamma$, a shift of $\Gamma \to \Gamma+ T$, where $T \in \Omega^1(X, T_X \otimes T_X^\ast)$, leads to  
\begin{align}
R &\to R + R + DT + T^2~,&
\tr R^2 & \to \tr R^2 + d \tr \left(2 T R + T DT + \ff{2}{3} T^3 \right)~.
\end{align}
While this causes no issues at the level of topology, it wreaks havoc on the complex structure decomposition of $\tr R^2$.  On a 3-fold we have
\begin{align}
\tr R^2 = (\tr R^2 )^{2,2} + (\tr R^2)^{3,1} + (\tr R^2)^{1,3}~.
\end{align}
On the other hand, $\tr \cF^2 $ is necessarily $(2,2)$, as is the left-hand side of~(\ref{eq:BIANCHI}).  For this reason it was suggested in~\cite{Strominger:1986uh} that one should use the Chern connection on $T_X$, which is guaranteed to produce a (1,1) curvature form.  

The connection is not arbitrary:  it is tied to a choice of field redefinitions involved in describing the higher order curvature corrections to supergravity.\footnote{A recent discussion in the context of (0,2) GLSMs was given in~\cite{Melnikov:2012nm}; it was based on earlier anomaly cancellation studies~\cite{Hull:1986xn,Sen:1986mg,Howe:1987nw}.}  From the world-sheet point of view we have two particularly nice choices dictated by the Green-Schwarz anomaly cancellation and its compatibility with world-sheet SUSY.  If we wish to preserve manifest (0,1) world-sheet supersymmetry, we can work with a covariant Hermitian metric $G$ on the space-time and then necessarily the Bianchi identity should be computed with the $\cS_+ = \Gamma + \ff{1}{2} H$ connection on $T_X$.  On the other hand, we can also try to preserve manifest (0,2) world-sheet SUSY.  This is possible and leads to the Chern connection in the Bianchi identity, but it requires that $G$ picks up non-trivial space-time Lorentz and gauge transformations of the same sort as familiar for the B-field.  This makes the geometry a bit more obscure to say the least!  Moreover, we expect that with such a choice the analysis of~\cite{Bergshoeff:1989de} will also be modified and produce $O(\alpha')$ shifts in the space-time SUSY conditions.

Hence, if we wish to work with a metric tensor and keep the space-time SUSY conditions in the form of~(\ref{eq:SUSY}), we must necessarily work with the $\cS_+$ connection.  What is then to be done about the (3,1) and (1,3) terms in the Bianchi identity?  A clue is provided by the Calabi-Yau case, where we have $\cS_+ = \Gamma_0 + O(\alpha')$, where $\Gamma_0$ is the K\"ahler (and in particular Chern) connection of the Calabi-Yau metric: while (1,3) and (3,1) terms are generated in the Bianchi identity, they are all $O(\alpha'^2)$.  We will now show that the resolution is similar in heterotic flux vacua, but with a twist.

\subsection{The torsional connection for heterotic fluxes}
Let us now consider $\cS_+$ for the heterotic flux solutions just described.  Since the Chern connection has particularly nice properties, it is useful to write $\cS_+ = \bSigma+\bT$, where $\bSigma$ is the Chern connection.  In terms of $6$-dimensional complex coordinates $dx^\mu$ $d\xb^{\mub}$, $\mu ,\mub=1,2,3$, we have
\begin{align}
\label{eq:SigT}
\bSigma & = 
\begin{pmatrix}
\Sigma^\mu_{\nu} & 0 \\
0 & \Sigmab^{\mub}_{\nub}
\end{pmatrix} =
\begin{pmatrix}
dz^\lambda g_{\nu\lambdab,\lambda} g^{\lambdab \mu} & 0 \\
0 & d\zb^{\lambdab} g_{\lambda\nub,\lambdab} g^{\mub \lambda}
\end{pmatrix}~,\nonumber\\
\bT & = 
\begin{pmatrix} 
0 & T^{\mu}_{\nub} \\ \Tb^{\mub}_{\nu} & 0 \end{pmatrix}
=
\begin{pmatrix}
0 & dz^\lambda H^{\mu}_{~~\lambda\nub} \\
d\zb^{\lambdab} H^{\mub}_{~~\lambdab\nu} & 0
\end{pmatrix}~.
\end{align}
Explicit computations turn out to be ugly in this basis, and rather than working with (1,0) forms $dz^i,d\theta$, we prefer to work with $dz^i,\Theta$.  This leads to simplifications.   For instance, the six-dimensional metric takes the form
\begin{align}
\bg = 
\begin{pmatrix} dz^k &d\theta \end{pmatrix}
\begin{pmatrix} 
g_{k\kb} + \ba A_k \Ab_{\kb} & \ba A_k \\ 
\ba\Ab_{\kb} & \ba \end{pmatrix}
\begin{pmatrix} d\zb^{\kb} \\ d\thetab \end{pmatrix}
=
\begin{pmatrix} dz^k &\Theta \end{pmatrix}
\begin{pmatrix} g_{k\kb} & 0 \\ 0 & \ba \end{pmatrix}
\begin{pmatrix} d\zb^{\kb} \\ \Thetab \end{pmatrix} ~.
\end{align}
Here $g_{k\kb}$ is the metric corresponding to the horizontal component of $\omega_X$, i.e. the metric on the base $M$ corresponding to Hermitian form $e^{2\vphi} \omega_0$, while $A=A^1+iA^2$ is the complexified connection $1$-form for the $T^2$ fibration.

It is not much more difficult to evaluate the resulting Chern connection and torsion.  We introduce a basis for $T_X$ dual to the $\{dz^i,\Theta\}$ basis for $T_X^\ast$ as follows: while the dual basis to $\{dz^i,d\theta\}$ is $\{\frac{\p}{\p z^i},\frac{\p}{\p\theta}\}$, the dual to $\{dz^i,\Theta\}$ is $\{E_i,E_\Theta\}$ with
\begin{align}
\begin{pmatrix} E_i \\ E_\Theta
\end{pmatrix}
=
\begin{pmatrix}
\delta^k_i & -A_i \\ 0 & 1 
\end{pmatrix}
\begin{pmatrix} \frac{\p}{\p z^k} \\ \frac{\p}{\p\theta}
\end{pmatrix},\qquad
\begin{pmatrix} \frac{\p}{\p z^i} \\ \frac{\p}{\p\theta}
\end{pmatrix}
=
\begin{pmatrix}
\delta^k_i & A_i \\ 0 & 1 
\end{pmatrix}
\begin{pmatrix} E_k \\ E_\Theta
\end{pmatrix}~.
\end{align}
With this choice of frame straightforward computation yields
\begin{align}
\Sigma 
& = 
\begin{pmatrix}
dz^p & \Theta 
\end{pmatrix} \otimes
\begin{pmatrix} dz^m \Gamma^q_{mp} & dz^m A_{p,m} \\ \ba dz^m \Fb_{m\jb} g^{\jb q} & 0 
\end{pmatrix} \otimes
\begin{pmatrix} E_q \\ E_\Theta
\end{pmatrix}~.
\end{align}
Here $dz^m \Gamma^q_{mp} = dz^m g_{p\pbar,m} g^{\pbar q} $ is the Chern connection on the base $M$ with respect to $g$.  The remaining term in the connection in~(\ref{eq:SigT}) is
\begin{align}
T & = 
\begin{pmatrix} d\zb^{\jb} & \Thetab \end{pmatrix} \otimes
\begin{pmatrix}  
dz^k H_{\nb k\jb}g^{\nb i} +\frac{\ba}{2} \Theta\Fb'_{\jb\nb} g^{\nb i} & 
-\frac{1}{2} dz^k F_{k\jb} \\
\frac{\ba}{2} dz^k F_{k\nb} g^{\nb i} & 0
\end{pmatrix} \otimes
\begin{pmatrix} E_i \\ E_{\Theta} \end{pmatrix}~.
\end{align}
Thus, we see that most of the terms in the connection are horizontal 1-forms, the only exception being
\begin{align}
\label{eq:Tproblem}
T \supset d\zb^{\jb} \otimes \left[\frac{\ba}{2}  \Theta \Fb'_{\jb\nb} g^{\nb i}\right]\otimes E_i~.
\end{align}
The full $R_+$ curvature is then
\begin{align}
\label{eq:Rplus}
R_+ = 
\begin{pmatrix} \pb \Sigma - \Tb T  &  \pb T -\Sigmab T  \\ \p\Tb-\Sigma \Tb & \p \Sigmab - T\Tb 
\end{pmatrix}
+
\begin{pmatrix} 0 & \p T -T  \Sigma  \\ \pb \Tb - \Sigmab\Tb & 0 
\end{pmatrix}.
\end{align}
While the first term is (1,1), the second contains (2,0) and (0,2) components.  So, in general  the curvature of the $\cS_+$ connection need not be (1,1), and  there are in general (3,1) and (1,3) components in $\tr R_+^2$.  

\subsection{Solving the Bianchi identity}
Taken naively,~(\ref{eq:Tproblem}) suggests that unless $\Fb' = 0$, i.e. we restrict to fluxes preserving N=2 space-time SUSY, there is simply no solution to the Bianchi identity.  Two observations make it clear that this is not the case:  first, we only know the Bianchi identity up to $O(\alpha'^2)$ corrections, and second, as we discussed above, $\ba$ is quantized in units of $\alpha'$.  Hence, to the order that we know the equation, we can neglect the terms proportional to $\ba$ in $\Sigma$ and $T$.  This leads to a purely horizontal connection, which has a (2,2) and horizontal $\tr R_+^2$.

Coming back to~(\ref{eq:fluxBianchi}), we see that we need to solve
\begin{align}
2i \p\pb e^{2\vphi} \omega_0 + \ba \p \Fb' \Theta + \ba \pb F' \Thetab & = \frac{\alpha'}{4} \left[ \tr R_+^2 - \tr \cF^2 + \ff{4\ba}{\alpha'} ( F \Fb - F' \Fb') \right] + O(\alpha'^2)~.
\end{align}
The right-hand side is a horizontal (2,2) form, whence we conclude that we need 
\begin{align}
\pb F' = 0 \implies F' = \lambda \Omega_0
\end{align}
for some constant $\lambda$.  Note that this also implies $\p F = 0$.   Finally, we solve the remaining equation in exactly the same manner as we obtained the $O(\alpha')$ correction to $T^2\times \text{K3}$ compactification above.  As long as anomaly cancellation is satisfied, 
\begin{align}
\frac{\alpha'}{4} \left[ \tr R_+^2 - \tr \cF^2 + \ff{4\ba}{\alpha'} ( F \Fb - F' \Fb') \right] = i \alpha' \p \pb f \omega_0~,
\end{align} 
we find an $O(\alpha')$ solution by setting $\vphi = \alpha' f/16$.  Note that $\vphi$ also appears in $\cS_+$, but this just gives an $O(\alpha'^2)$ contribution to the Bianchi identity that can be ignored to this order.

\subsubsection*{A comment on the Bianchi identity for N=2 solutions}
So far what we have said applies equally well to solutions with both N=1 and N=2 space-time SUSY.  However, in the latter case we observe that the connection $\cS_+$ is purely horizontal without any $\alpha'$ expansion, and hence $\tr R_+^2$ is (2,2) and horizontal.  In this case one may attempt to solve the Bianchi identity for $\vphi$ without any $\alpha'$ expansion.  This leads to a non-linear PDE for $\vphi$, and it is not a priori clear that a smooth solution exists.  A similar equation, where $\tr R^2$ is evaluated for the Chern connection, was studied in~\cite{Fu:2006vj,Becker:2006et} and shown to possess the requisite solutions.  Examining the proof given in~\cite{Fu:2006vj} it is clear the the result is easily modified to apply to the equation with the $\cS_+$ connection:  the details of the connection choice are all incorporated into a single function $g$ that appears in the PDE for $\vphi$, and $g$ is not required to have any special properties besides being $C^{\infty}$.

It would be interesting to understand whether  this full non-linear solution has any physical significance.  In the most optimistic case it may be that for N=2 SUSY backgrounds, the higher order $\alpha'$ corrections to the Bianchi identity and SUSY variations vanish, so that the leading order non-linear solution is in fact exact.  If the higher order corrections do not vanish, perhaps it is possible to show that they can all be incorporated into a correction of the function $g$.  Such a result would be akin to~\cite{Nemeschansky:1986yx}, which shows that the Calabi-Yau metric can be corrected order by order to solve the all orders beta function for a (2,2) NLSM with Calabi-Yau target space.  For heterotic flux backgrounds with N=2 space-time supersymmetry such a program may be feasible.

\subsubsection*{HYM for $R_+$}
There is one more observation we would like to make.  A number of works have emphasized that in order for a configuration satisfying the SUSY conditions and Bianchi identity to be a solution to the heterotic equations of motion to $O(\alpha')$, the curvature $R_+$ appearing in the Bianchi identity must satisfy HYM equations on $X$ up to $O(\alpha')$ corrections~\cite{Ivanov:2009rh,Fernandez:2014kwa}.  In the case of Calabi-Yau compactification this is obviously satisfied, since $R_+ = R_0 + O(\alpha')$, where $R_0$ is the Calabi-Yau curvature 2-form.  In case of the heterotic flux vacua we considered, this condition is obeyed as well.  To see this, note that since $\ba$ is quantized and $\vphi$ is also of order $\alpha'$, to leading order
\begin{align}
R_+^{2,0} & = \begin{pmatrix} 0 & \p T \\ 0 & 0 \end{pmatrix} + O(\alpha') ~&
T & = d\zb^{\jb} \otimes\left[-\ff{1}{2} dz^{k} F_{k\jb} \right] \otimes E_{\Theta} + O(\alpha')~.
\end{align}
Since $\p F = 0$, we see that $R_+^{2,0} = O(\alpha')$.  We can also check that 
\begin{align}
\omega_X^2 R^{1,1}_+ = \frac{i\ba}{2} \Theta\Thetab \omega_0 R^{1,1}_+ = O(\alpha'^2)~.
\end{align}
To see this, we work out the various components of $R_+^{1,1}$ and find that up to $O(\alpha')$ $R^{1,1}_+$ is horizontal, and every component is proportional to either $\Rh$, the curvature of the base Ricci-flat metric $\gh$ on the base or a $\gh$-covariant derivative of $F$.  All these terms are annihilated by $\omega_0$, and the result follows.

\section{The (dis)connection between N=2 and N=1 flux vacua}
A well-known fact about Minkowski SUSY string vacua is that motion on the moduli space preserves space-time supersymmetry.  In perturbative heterotic string theory this has been made very concrete~\cite{Banks:1988yz,Dixon:1987bg}:    an N=1 SUSY vacuum corresponds to a  (0,2) SCFT with integral $\GUR$ charges; every marginal heterotic deformation must preserve (0,1) superconformal invariance since this is a left-over gauge symmetry of the heterotic string, and it is easy to show that every (0,1) marginal deformation of a (0,2) SCFT with integral $\GUR$ charges preserves the full (0,2) superconformal invariance.   This is easily extended to perturbative heterotic vacua with N=2 SUSY.~  In this case the $\cb = 9$ right-moving $N_{\text{ws}}=1$ superconformal algebra\footnote{The subscript ``ws'' is used to lessen confusion between space-time SUSY and the right-moving world-sheet superconformal algebras.} (SCA) splits as a sum of a $\cb = 3$ $N_{\text{ws}}=2$ and $\cb = 6$ $N_{\text{ws}}=4$ SCAs, and any $N_{\text{ws}}=1$ marginal deformation preserves the split into $N_{\text{ws}}=2$ and $N_{\text{ws}}=4$ SCAs.\footnote{The proof uses the result of~\cite{Banks:1988yz} that a diagonal $\cb = 9$ $N_{\text{ws}}=2$ SCA is necessarily preserved, and one can show that $N_{\text{ws}} = 2$ marginal deformations must preserve the $\SU(2)_{\text{R}}$ Kac-Moody algebra of the $N_{\text{ws}}=4$ theory.  From that it easily follows that all marginal deformations preserve the full $N_{\text{ws}}=2$ and $N_{\text{ws}}=4$ SCAs and hence the full space-time SUSY.}  This is familiar from the space-time point of view:  partial SUSY breaking cannot be achieved by motion on the moduli space; rather it requires deforming the SUSY current algebra and hence the UV physics.

This simple point seems to lead to a small paradox for the flux solutions described above.  We saw that the distinction between N=2 and N=1 solutions has to do with whether the complex curvature 2-form $\bF = \bF^1 + i \bF^2$ has a (2,0) component on the base K3, and we also saw that this (2,0) component takes the form $F' = \lambda \Omega_0$ for some complex constant $\lambda$.  It is well-known that the decomposition of a 2-form on a K3 surface $M$ is not a topological condition:  small changes in the complex structure of $M$ can produce small (2,0) and (0,2) components of a (1,1) form in the original complex structure.  Naively, this suggests that $\lambda$ is a continuous parameter that interpolates between N=2 ($\lambda=0$) and N=1 solutions.

\subsubsection*{Dual M-theory perspective}
To resolve this conundrum, we first consider the dual perspective, where we build a $d=3$ compactification of M-theory on a product of two K3s, $Y = M_1 \times M_2$.  To find such solutions we must turn on a non-trivial $G$-flux on $Y$.  Anomaly cancellation requires
\begin{align}
\frac{1}{2} \int_Y \frac{G}{2\pi} \frac{G}{2\pi} = \frac{\chi(Y)}{24} = 24~
\end{align}
with $\ff{1}{2\pi} G \in H^4(Y,\Z)$, and supersymmetry requires that it is (2,2) and primitive with respect to the K\"ahler form $\omega_Y = \omega_1 + \omega_2$~\cite{Becker:1996gj, Dasgupta:1999ss}.  In general, a solution of this sort will break the hyper-K\"ahler symmetries of $M_{1,2}$ which act on $\omega_{1,2}, \Omega_{1,2},\Omegab_{1,2}$.  On the other hand, a flux $G$ compatible with the hyper-K\"ahler symmetries, i.e. one satisfying
\begin{align}
\label{eq:strongflux}
\omega_{1,2}  G &= 0~,&
\Omega_{1,2} G & = 0~,&
\Omegab_{1,2} G & = 0~
\end{align}
is necessarily (2,2) and primitive on $Y$.  This flux preserves $8$ supercharges in three dimensions, while the more generic choice only preserves $4$ supercharges~\cite{Dasgupta:1999ss}.

In this case again, one might naively think that infinitesimal deformations of complex structure of $M_1$ and $M_2$ could possibly transform the special form of $G$-flux to a less generic one.  Here it is easy to see that this is not the case.
For simplicity, consider a flux of the form 
\begin{align}
\ff{1}{2\pi } G =  \xi_1 \xi_2~,
\end{align}
where $\xi_{1,2}$ are integral anti-self-dual classes in $H^{2}(M_{1,2},\Z)$.  Evidently such a flux satisfies~(\ref{eq:strongflux}).  Under a small perturbation of complex structure of $M_1$ and $M_2$ we have
\begin{align}
\xi_1 = \eta_1 + \ep_1 \omega_1 + \lambda_1 \Omega_1 + \lambdab_1 \Omegab_1~,
\end{align}
where $\ep_1$ and $\lambda_1$ are infinitesimal parameters and $\eta_1$ is anti-self-dual and similarly for $\xi_2$.  Plugging this into $G$, we find that its (3,1) component is given by
\begin{align}
\ff{1}{2\pi} G^{3,1} = \delta_1\Omega_1 (\eta_2+\ep_2\omega_2) + (\eta_1+\ep_1\omega_1) \delta_2 \Omega_2~.
\end{align}
We see that this vanishes if and only if $\delta_1 = \delta_2 =0$ --- no cancellation is possible between the two factors.
Similarly, primitivity with respect to $\omega_Y$ requires $\ep_1 = \ep_2 = 0$.  Hence, for all allowed infinitesimal deformations the flux satisfies~(\ref{eq:strongflux}).  This is easily generalized to any deformations of the most general flux satisfying~(\ref{eq:strongflux}).  A deformation with non-zero $\delta$ or $\ep$ necessarily breaks all supersymmetry and will have a non-trivial potential.

Thus, in this dual frame it is clear that the solutions do have the expected behavior:  infinitesimal deformations preserve the space-time SUSY.  This has two important lessons for understanding the ``puzzle'' in the heterotic frame.  First, the conundrum is resolved at the level of the eleven-dimensional SUSY conditions on the flux; second, it is the interplay between the two K3s that eliminates the paradoxical deformations.  In the heterotic frame this strongly suggests that the resolution also involves the ten-dimensional SUSY conditions, and in particular the interplay between the base K3 geometry and the $3$-form flux.  As we will now show, that is the case.

\subsubsection*{Deformations of complex structure and $H$}
Returning to the heterotic frame, we consider an N=2 space-time solution, where $\omega_0 \bF = \Omega_0 \bF = 0$, and we examine the form of the $H$ flux as we deform the complex structure of the base K3.  
 Let $s \in H^1(M,T_M)$ be an infinitesimal complex structure deformation on $M$.  This lifts to an infinitesimal complex structure deformation on the full six-dimensional $X$ if and only if 
\begin{align}
s \llcorner \bF = \ff{1}{2} (s^i_{\jb} \bF_{i\kb} - s^i_{\kb} \bF_{i\jb}) d\zb^{\jb} d\zb^{\kb} = 0~. 
\end{align}
This is just the condition that under the deformation $\bF$ does not acquire a (0,2) component.  In this case we can make a gauge transformation  so that the $T^2$ fibration connection $A^1 + iA^2$ is a local (1,0) form and $\Theta$ is a global (1,0) form on $Y$ in the new complex structure.

Using such an $s$ we construct the projectors $\Pi^{p,q}_s$ onto $\Omega^{p,q}(M)$ in the deformed complex structure.  With these, we have
\begin{align}
\bF = \Pi_0^{1,1} \bF = \Pi_s^{1,1} \bF + \Pi_s^{2,0} \bF~.
\end{align}
The $3$-form flux for the undeformed N=2 solution then takes the form
\begin{align}
H = H_{\text{hor}} -\frac{\ba}{2} (\Pi^{1,1}_0 \bF) \Thetab - \frac{\ba}{2} (\Pi^{1,1}_0 \bFb) \Theta~.
\end{align}
Under the deformation we then have
\begin{align}
H_{\text{def}} = (H_{\text{hor}})_{\text{def}} - \frac{\ba}{2} ( \Pi^{1,1}_s \bF + \Pi^{2,0}_s \bF) \Thetab -\frac{\ba}{2} (\Pi^{1,1}_s \bFb + \Pi^{0,2}_s \bFb) \Theta.
\end{align}
Comparing this with the form of flux required to preserve N=1 space-time supersymmetry, we find a crucial difference in the signs of the (2,0) and (0,2) contributions:
\begin{align}
H_{N=1} = H_{\text{hor}} - \frac{\ba}{2}  (\Pi_s^{1,1} \bF-\Pi^{2,0}_s \bF ) \Thetab - \frac{\ba}{2} (\Pi_s^{1,1} \bFb-\Pi^{0,2}_s\bFb) \Theta~.
\end{align}
We therefore see that a variation of complex structure of the N=2 solution either breaks or preserves all supersymmetry.  In the former case we expect a space-time potential will ensure that the deformation is not in fact marginal.  This difference in signs again underscores the rather different roles played by the (1,1) and (2,0) components of the $T^2$ curvature 2-form as already observed above in~(\ref{eq:brokenKM}).

\section{Discussion} \label{s:disc}
We have seen that, as far as supergravity is concerned, if we are careful about the consequences of flux quantization, there is no trouble in solving the SUSY and Bianchi conditions.  This success must be taken with a large grain of salt:  the fact that $\ba$ is quantized means that neglecting the higher $\alpha'$ corrections is a rather formal exercise.  Indeed, from this perspective it is remarkable that the $\alpha'$ expansion seems to give any sensible results at all.  For the case at hand we had a number of additional tools that may not be available in general.  First, while these flux solutions do not have a ten-dimensional large radius limit, they do have an eight-dimensional large radius limit, and the fact that we have a  good understanding of heterotic compactification on $T^2$ allows us to discuss the solutions from that point of view.  Second, we have a class of solutions that preserve N=2 space-time supersymmetry, so that if they exist, any quantum corrections (whether in $\alpha'$ or $g_s$) are severely constrained.
Finally, we have a rather explicit dual description in terms of F/M-theory compactification on $\text{K3}\times\text{K3}$, which again strongly suggests that these solutions should exist as bona fide vacua of string theory.  As for Mallarm{\'e}'s \textit{ptyx}~\cite{Kromer:ptyx}, while some properties of the N=2 flux solutions are obscure, there is little doubt that they exist. 

It would be interesting to determine to what extent these sorts of techniques are applicable to more general  N=1 heterotic flux vacua.  That is difficult to establish given the lack of concrete examples of admissible topologies and complex geometries.  It would be instructive to have a large class of admissible topologies, ideally something akin to the Kreuzer-Skarke list of Calabi-Yau hypersurfaces in toric varieties.

Given that a solution exists, the next step is to identify its basic properties.  For N=2 solutions the special properties just mentioned give a great deal of control.  For instance, it is easy to describe the full massless spectrum and identify various branches of the moduli space via the space-time super-Higgs mechanism.  For instance, many examples were examined at this level in~\cite{Melnikov:2012cv}.  Moreover, one can also find dual IIA compactifications on K3-fibered Calabi-Yau three-folds, and it should be in principle possible to use the IIA and heterotic descriptions to actually determine the full quantum-corrected metric on the moduli space.\footnote{This is by no means an easy task to carry out in practice!  See~\cite{Alexandrov:2013yva} for a recent review.} 

More generally, for N=1 flux backgrounds even identifying the massless spectrum is a great challenge even at the formal level of $\alpha'$-corrected supergravity.  Recent progress in this direction has been made in~\cite{Melnikov:2011ez,Anderson:2014xha,delaOssa:2014cia}, but it rather begs the question of why an $\alpha'$ expansion should be at all sensible in these more general backgrounds.  As we noted above, both in the Calabi-Yau and heterotic flux case it turns out to be possible to find all order solutions to the $O(\alpha')$ SUSY and Bianchi equations.  It would be interesting to understand whether these solutions have any physical significance, and a detailed study of $O(\alpha'^2)$ corrections may be illuminating.

Finally, if we can make sense of the all orders $\alpha'$ expansion, then we will need to consider non-perturbative corrections in $\alpha'$.  It is known in Calabi-Yau compactifications that world-sheet instantons do contribute to space-time potentials and modify spectra, but it is not known how to make sense of such configurations in the presence of $H$-flux, where the instanton action is necessarily complex, and the imaginary contribution is not topological.  It is likely that one must examine more general saddle points in a suitably continued field space of the NLSM.  Making sense of an expansion in terms of such saddle points would be a very significant development in extracting quantitative space-time physics from heterotic geometries.

\bibliographystyle{./utphys}
\bibliography{./bigref}

\end{document}